# Simulations on a potential hybrid and compact attosecond X-ray source based on RF and THz technologies

T. Vinatier[a,1], R. W. Assmann[a], U. Dorda[a], F. Lemery[a], B. Marchetti[a]
[a]Deutsches Elektronen Synchrotron, Notkestrasse 85, 22607 Hamburg, Germany
[1]Corresponding author: thomas.vinatier@desy.de

*Abstract*
We investigate through beam dynamics simulations the potential of a hybrid layout mixing RF and THz technologies to be a compact X-ray source based on Inverse Compton Scattering (ICS), delivering few femtoseconds to sub-femtosecond pulses. The layout consists of an S-band gun as electron source and a dielectric-loaded circular waveguide driven by a multicycle THz pulse to accelerate and longitudinally compress the bunch, which will then be used to produce X-ray pulses via ICS with an infrared laser pulse. The beam dynamics simulations we performed, from the photocathode up to the ICS point, allows to have an insight in several important physical effects for the proposed scheme and also in the influence on the achievable bunch properties of various parameters of the accelerating and transverse focusing devices. The study presented in this paper leads to a preliminary layout and set of parameters able to deliver at the ICS point, according to our simulations, ultrashort bunches (≈ 1 fs rms), at 15 MeV, with ≥ 1 pC charge and transversely focused down to ≈ 10 µm rms while keeping a compact beamline (≤ 1.5 m), which has not yet been achieved using only conventional RF technologies. Future studies will be devoted to the investigation of several potential ways to improve the achieved bunch properties, to overcome the limitations identified in the current study and to the definition of the technical requirements. This will lead to an updated layout and set of parameters.

## 1. Introduction

Particle acceleration beyond the few-MeV level currently requires large infrastructures, of the order of several meters or tens of meters, due to the fact that the conventional RF accelerating structures operate at low frequencies (a few GHz) and with relatively low field amplitudes (a few tens of MV/m in the meter-long structures). The same remark holds for the schemes used to compress electron bunches down to lengths of the single femtosecond order or below. This is indeed typically done via the velocity bunching process [1] in several meters long traveling wave accelerating structures or in magnetic chicanes, which length depends on the bunch energy but is typically at least of several meters. Compression down to a few femtoseconds or below can also be achieved by using a short (≈ 10 cm long) standing wave buncher placed right after an RF-gun, like it is done on the REGAE facility [2, 3], but in this case the bunch charge remains limited to values well below 1 pC. In this study, we aim to simulate the possibility of delivering such short bunches with at least 1 pC of charge.

One of the schemes currently studied to address the two aforementioned issues and significantly reduce the footprint of particle accelerators is to use dielectric-loaded structures driven by laser-generated THz fields [4, 5], in which the frequencies (100 GHz to 10 THz) and field amplitudes (100 MV/m to a few GV/m) are expected to be much higher than in conventional RF accelerating structures. This would allow bunch acceleration and compression by velocity bunching within a few tens of cm, thus reducing the footprint of accelerator beamlines by up to two orders of magnitude compared to conventional RF technologies. This is what the AXSIS project [6] aims for. In this project, the electron bunch is intended to be photo-emitted and accelerated in a gun driven by single-cycle THz pulses [7] up to an energy of the order of 1 MeV. It will be then accelerated up to 15-20 MeV and compressed down to 1 fs rms or below in a linac structure consisting in a hollow dielectric-loaded circular waveguide driven by a multi-cycle THz pulse [8, 9, 10] exciting the $TM_{01}$ mode [11, 12]. This structure will be named THz linac hereafter. The goal is to obtain a 1 pC and 15-20 MeV electron bunch to produce attosecond X-ray pulses by Inverse Compton Scattering (ICS) [13, 14] with an infrared laser pulse after transverse focusing.

The all-optical scheme intended for AXSIS has intrinsic desirable features: the compactness, the possibility to run at higher repetition rates than with conventional RF technologies due to the absence of high power RF pulses, and also the fact that such an accelerator would be ideally self-synchronized since all would come from one unique source laser (cathode laser, THz accelerating fields and laser for ICS). However, the last point is true only if the amplitude of the source laser is sufficiently stable from one shot to the next. Indeed, fluctuations of the source laser amplitude will lead to variations in the electron bunch energy and therefore of its travel time along the beamline. This could result in a significant desynchronization between the electron bunch and the various fields driven by the source laser. The level of stability required for AXSIS is not yet defined.

Despite these potential advantages of using THz guns, we choose to investigate in this paper an alternative scheme to the AXSIS baseline design aforementioned for a compact attosecond X-ray source based on ICS. This alternative scheme consists in replacing the foreseen THz gun [6, 7] by a conventional S-band RF-gun. Subsequently, the generated electron bunch will be injected in a THz linac similar to the one foreseen in the AXSIS project [15]. In the following paragraphs, we will detail the reasons motivating this choice.

First of all, while the THz guns are still in the early phase of their development [16, 17, 18], the S-band guns are well-known, characterized and routinely operated electron sources. Besides they are currently the best sources to provide ultrashort and high-brightness electron bunches, making them a natural starting point for a compact attosecond X-ray source based on ICS. Furthermore, when dealing with new and developing technologies, it is preferable to introduce only one new part at a time (the THz linac in the case simulated in this paper) and to keep the other components conservative and well-established. In addition to these reasons, several physical arguments also explain the choice of replacing the THz gun with an S-band RF-gun.

First, the THz guns aim to work in a regime never explored by the other electron sources. This can be visualized in Fig. 1, which represents the *α* parameter as a function of the field frequency for various electron sources. The normalized accelerating field strength $\alpha = eE_0/2m_ec^2k$ [19] is often encountered in the literature about RF-guns [20, 21]. Here $e$ is the elementary charge, $E_0$ the accelerating field amplitude, $m_e$ the electron mass, $c$ the speed of light in vacuum and $k$ the wave vector of the accelerating field. *α* tells how quickly the bunch will be accelerated and shows relativistic effects compared to the field wavelength and entirely determines the longitudinal dynamics in the gun. It is well below 1 for THz gun designs and therefore very low compared to the other electron sources, namely RF-guns and plasma injectors, where $\alpha$ is higher or close to 1 (see Fig. 1). This leads to an intrinsic uncertainty in the quality of the bunch which will be obtained with such devices. This low value of $\alpha$ is explained by the fact that while the frequency is two to three orders of magnitude higher than for the RF-guns, the expected field amplitude in THz guns is only about one order of magnitude higher at maximum (see the design shown in [7]) and therefore remains well below the 6.5 GV/m which would be needed to reach $\alpha = 1$ at 300 GHz. Note that the field amplitude can effectively be two to three orders of magnitude higher than for RF-guns in case of plasma injectors, allowing $\alpha$ to approach 1. The current limitation in the field amplitude for the THz guns is due to difficulty in producing a sufficient amount of THz power.

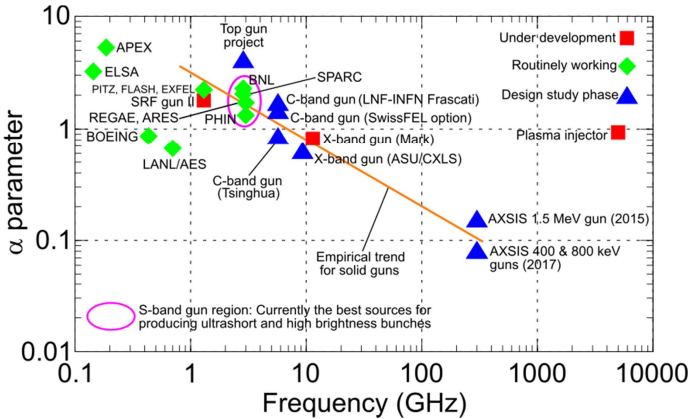

**Fig. 1.** $\alpha$ parameter as a function of the operating frequency for several types of electron sources.

Moreover, some aspects of the development of the THz guns are challenging and raised issues currently addressed by the AXSIS collaboration. First, the limited transverse aperture of this type of gun (120 µm diameter for the design shown in [7]) induces significant charge losses due to space-charge forces and to the effect of the off-axis transverse fields. This makes the extraction of a significant charge ($\geq 1$ pC) from this type of gun challenging, unless a strong magnetic field of typically several Teslas is used to confine the electron bunch during its path in the gun. The application of this field on the photocathode during the bunch emission is however questionable, since it will strongly deteriorate its emittance and couple the dynamics between the two transverse planes, making it more difficult to transport. Then, the UV laser pulses currently realistically available to generate electron bunches by photoemission have a minimal duration typically around 30 fs rms. This duration is not short compared to the typical period of the accelerating field in the THz guns (for the case shown in [7], 1° of phase corresponds to 9.3 fs). The first consequence of this condition is that the bunch energy spread generated in a THz gun will be much higher than in a conventional RF-gun. The second consequence, of particular relevance for the generation of ultrashort bunches, is that more important curvature and other non-linearities will be generated in the bunch longitudinal phase-space during its acceleration in a THz gun than in a conventional microwave RF-gun. These non-linearities will strongly limit the minimal bunch length achievable subsequently by compression by velocity bunching in a THz linac.

One objective of our simulation study is to characterize the capabilities of the THz linac foreseen in the AXSIS project for reaching the requirements in terms of bunch properties at the ICS point, which are shown in Table 1. One should note that this includes the final focusing to the ICS point located after the THz linac. The study that we present in this paper has been conducted in such a way that it is also a first investigation on a dedicated hybrid layout mixing RF and THz technologies for a compact attosecond X-ray source based on ICS. This hybrid scheme is of interest since it would allow straightforwardly overcoming the aforementioned difficulties arising with the use of THz guns, while still keeping the advantages of the THz linac over the conventional RF technologies (compactness and efficiency of bunch compression by velocity bunching). Note that this scheme would require a time synchronization between the electron bunch and the THz pulse driving the linac at the 10 fs level or below (1° of phase ≡ 9.3 fs at 300 GHz). This level has been demonstrated at REGAE [22] and is intended on the ARES linac [23]. We will investigate in this paper, through simulations, the capability of this hybrid scheme to generate bunches fulfilling the AXSIS requirements (see Table 1) at a potential ICS point, while keeping a compact beamline ($\leq 2$ m) which has, at the time of writing, not been achieved using only conventional RF technologies.

**Table 1:** Requirements for bunch properties at the ICS point in the AXSIS project.

| Bunch charge | Bunch mean kinetic energy | Rms bunch length | Rms bunch transverse dimension |
|---|---|---|---|
| $\geq 1$ pC | 15-20 MeV | $\leq 1$ fs | $\approx 10$ µm |

We will only address the aspect of beam dynamics simulations up to the ICS point, and not the photon production by the ICS process and the technical feasibility of the scheme (especially the requirements in terms of beam diagnostics and steering and in terms of properties of the THz pulse driving the THz linac). These topics will be included in future studies. Another ongoing study within the AXSIS collaboration is investigating the tolerances for beam dynamics in THz-driven dielectric-loaded circular waveguides with respect to various fabrication errors and experimental jitters [24].

In this paper, we will present ASTRA simulations [25] using several layouts based on the introduced hybrid scheme, a schematic of the general layout being shown in Fig. 2, with the goal to generate electron bunches with properties suitable for producing attosecond X-ray pulses via ICS (see Table 1). In the first section, we will concentrate on the part between the photocathode and the entrance of the THz linac, looking especially at the impact of the transverse size of the UV laser pulse generating the electron bunch in the gun and of the solenoid used to focus the bunch into the THz linac. The acceleration and compression of the bunch in the THz linac will be developed in the second section, paying particular attention to the effect of the frequency of the accelerating field, of the interaction length between the electron bunch and the THz accelerating field and of the distance between the photocathode and the THz linac entrance. Then, in the third section, the transverse focusing of the bunch to the potential ICS point will be investigated and the achieved bunch properties at this point presented. Because of the relatively low energy intended in our scheme, special care will be taken that the transverse and longitudinal focal points of the bunch coincide with each other at the potential ICS point. Finally, we will also expose the limitations of the scheme we investigated from the beam dynamics point of view.

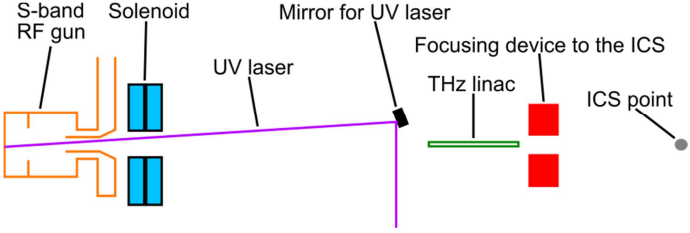

**Fig. 2.** Schematic of the general layout for the hybrid scheme proposed in this paper.

## 2. From the photocathode to the THz linac entrance

*2.1. Assumptions and conditions*

For the study from the photocathode up to the entrance of the THz linac, we fix several general parameters. In all the following studies we consider as electron source a 1.6 cell S-band gun operating at 2.9985 GHz, taking as a reference the gun intended for the ARES linac at SINBAD [23]. Then, for injecting the electron bunch into the THz linac, we choose a solenoid having two coils with reversed polarities and the center located 40 cm away from the photocathode. We also assume that the UV laser driving the RF-gun has a 3D Gaussian shape.

The main goal of our study is to produce an electron bunch with an rms length $\sigma_t$ of the single femtosecond order or below after compression by velocity bunching in the THz linac. Apart from the space-charge forces, one of the main limitations is the curvature induced by the high-frequency field on the bunch longitudinal phase-space in the THz linac during compression. To minimize it, it is important that the bunch enters the THz linac with the shortest possible length. To fulfil this goal we choose to fix the rms length of the UV laser pulse at the cathode $\sigma_{tL}$ to 75 fs, which corresponds to the minimum achievable on the ARES linac [23]. Then the gun RF-phase $\varphi_0$ is chosen to provide the maximum of longitudinal compression at the photocathode (which happens when the electric field is zero and starts to become accelerating) taking care that all the bunch charge is effectively extracted from the cathode. These choices allow minimizing $\sigma_t$ at the THz linac entrance while not inducing strong non-linearities of the longitudinal phase-space, which would otherwise limit the achievable bunch length by velocity bunching in the THz linac. We will also limit the study to the two relatively low bunch charges $Q = 1$ pC and $Q = 5$ pC, which are still significantly higher than the ones currently used to produce femtosecond bunches with a short buncher located at the exit of an RF-gun [2, 3]. Finally, the two relatively high field amplitudes in the gun $E_{mg} = 110$ MV/m and 140 MV/m have been simulated, which correspond respectively to the one expected at ARES [23] and to the maximal achievable with a BNL/SLAC/UCLA gun [26, 27].

The space-charge forces tend to increase the bunch length in the drift space between the gun and the THz linac entrance. To minimize this effect, and have the shortest possible bunch length at the THz linac entrance, the length of this drift space should be minimized. This imposes to escape the classical design schemes of RF photo-injectors, where the typical distance between the gun and the first accelerating section is $\geq$ 1.5 m. This will potentially raise issues for the mechanical integration of the beamline (vacuum system for example). The diagnostics would also have to be reduced to compatible sizes and numbers (charge, imaging and steering being a minimum). We have investigated two possible options for the position of the THz linac entrance $Z_{linac}$: One relatively close (around 60 cm away from the photocathode) and the other a little farther (around 85 cm away from the photocathode). This imposes restrictions on the type of RF-power coupling which could be used for the gun. Indeed, the case with a THz linac close to the photocathode requires a RF-power coupling on the side of the gun. The reason is that in the case of an on-axis coupling, like the one represented in Fig. 2, the aperture at the RF-coupler is reduced to around 15 mm diameter compared to the typical 25-30 mm diameter of the irises. The mirror for injecting the UV laser pulse into the gun (located before the THz linac, see Fig. 2) would in this case have to be too close from the axis and not enough margins will be provided for the passage of the electron bunch. For the case with a THz linac farther away from the photocathode, this constraint is relaxed and an on-axis RF power coupling could also be used for the gun.

Finally, the last thing we fix is the transverse dimension to which the bunch has to be focused by the solenoids into the THz linac. In Sec. 3, we will consider the two possible operating frequencies of 150 GHz and 300 GHz for the THz linac. We will assume that for 300 GHz the THz linac will have a 1 mm diameter of the vacuum channel. The thickness of the dielectric-loading in the linac will then depend on the material and on the required phase velocity $v_{ph}$ of the THz pulse exciting the $TM_{01}$ mode in the linac [11]. For $v_{ph} = c$, as assumed in this paper, and a quartz-loading ($\varepsilon_r = 4.41$), the thickness of the loading should be $\approx$ 90 µm. For 150 GHz, we will assume a 2 mm diameter vacuum channel. By defining the requirement on the bunch rms transverse dimension at the THz linac entrance to be 10 times smaller than the vacuum channel diameter, we fix it to 100 µm for 300 GHz and 200 µm for 150 GHz.

*2.2. Influence of the transverse dimension of the UV laser pulse and of the solenoid*

Once $Q$, $\sigma_{tL}$, $E_{mg}$ and $\varphi_0$ have been fixed in the way explained in Sec. 2.1, the two parameters driving the achievable $\sigma_t$ at the entrance of the THz linac are, in a coupled way, the rms transverse dimension of the UV laser pulse on the cathode $\sigma_{rL}$ and the solenoid strength used to focus the bunch. Note that the solenoid strength is not a free parameter, since it is fixed by the position of the THz linac entrance and the transverse dimension to which we want to focus the bunch both defined at the end of Sec. 2.1, but its effect depends on the bunch transverse size as will be shown later.

The achievable $\sigma_t$ at the entrance of the THz linac always presents, for the parameters we have chosen, a minimum for a certain value of $\sigma_{rL}$, denoted $\sigma_{rLm}$ thereafter, which we will use in our simulations. An example is shown on Fig. 3 (left) for one set of parameters, where $\sigma_t$ right at the exit of the RF-gun is also plotted. The presence of the minimum can be understood by looking at Fig. 3 (right), which shows the evolution of $\sigma_t$ from the cathode to the THz linac entrance for 3 different values of $\sigma_{rL}$ extracted from Fig. 3 (left).

The increase of $\sigma_t$ for $\sigma_{rL} < \sigma_{rLm}$ is due to space-charge forces which prevents the bunch from being compressed in the

RF-gun, especially right after the cathode where the energy is the lowest. This is visible on Fig. 3 (right) where $\sigma_t$ right after the cathode is much higher for $\sigma_{rL} = 0.2$ mm than for $\sigma_{rLm} = 0.7$ mm, and it is even higher than $\sigma_{tL}$ (75 fs rms). For $\sigma_{rL} > \sigma_{rLm}$, the space-charge forces immediately after the cathode are lower and the bunch is therefore more compressed. This is visible on Fig. 3 (right) where $\sigma_t$ right after the cathode is smaller for $\sigma_{rL} = 1.2$ mm than for $\sigma_{rLm} = 0.7$ mm. However during the rest of the path in the RF-gun, $\sigma_t$ exhibits a faster increase and finally exceeds the one for $\sigma_{rLm}$ at the gun exit. This is not due to an increase of space-charge forces induced by the stronger compression at the cathode, since the same behavior is observed when they are switched off in ASTRA. This is also not due to an over-compression of the bunch at the cathode, since the same behavior is observed when starting with no space-charge forces and $\sigma_{tL} = 0$ fs in ASTRA. This is actually due to the radial dependency of the electric field in the RF-gun, namely the decrease of the longitudinal component and the increase of the transverse one with increasing values of the radial offset from the RF-gun axis. This radial dependency induces a correlation between the transverse position and the longitudinal momentum of the electrons in the bunch. This correlation leads to an increase of $\sigma_t$ in the gun and in the drift space located after the gun, because the bunch is not yet ultra-relativistic and differences in longitudinal momentum translate in non-negligible velocity differences. This effect becomes obviously stronger when $\sigma_{rL}$ increases, because the longitudinal momentum (and therefore velocity) difference between on-axis and off-axis electrons becomes greater.

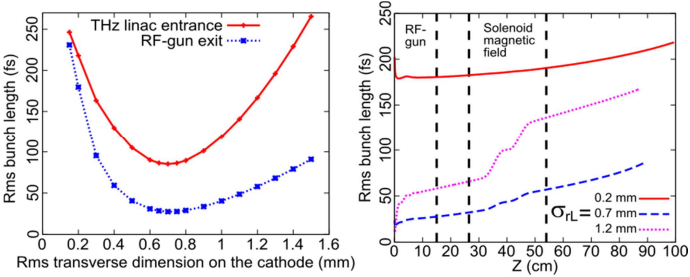

**Fig. 3.** $\sigma_t$ at the THz linac entrance and RF-gun exit as a function of $\sigma_{rL}$ (left) and evolution of $\sigma_t$ from the cathode up to the THz linac entrance for 3 values of $\sigma_{rL}$ (right). Conditions: $E_{mg}$ = 110 MV/m; Q = 5 pC; $\sigma_{tL}$ = 75 fs; $Z_{linac}$ = 89 cm and bunch transversely focused to 100 µm rms at this point.

Fig. 3 (left) also shows that the value of $\sigma_{rLm}$ is the same either we record $\sigma_t$ right at the gun exit or at the THz linac entrance. It demonstrates that the minimization of $\sigma_t$ at the THz linac entrance is only driven by $\sigma_{rL}$ and not by the solenoid. However, it is visible on Fig. 3 that the solenoid has an effect on the value of $\sigma_t$ at the THz linac entrance and that this effect becomes stronger when $\sigma_{rL}$ increases. One contribution is located inside the region where the solenoid magnetic field is located (see Fig. 3 (right)). It is due to the fact that the electrons with a larger transverse offset have a longer path in the solenoid magnetic field and they moreover travel more slowly, since in our case the correlation is such that the electrons with the largest transverse offset have the lowest longitudinal momentum. The result is an increase of the bunch length (similar to its decrease by magnetic compression in a chicane). The other contribution appears after exiting the solenoid magnetic field. It is simply explained by the fact that electrons with a larger transverse offset acquire a larger divergence and therefore travels a longer path to reach the same longitudinal position as electrons with smaller transverse offset. This leads to an increase of $\sigma_t$, amplified by the fact that these electrons with larger transverse offsets also have lower velocities. Such an effect of the divergence on $\sigma_t$ is already present before the solenoid (see Fig. 3 (right)), created in the RF-gun by the space-charge forces and RF transverse forces. But, in our case, this effect is smaller than the one of the solenoid-induced divergence. This is visible on Fig. 3 (right), where the rate of increase of $\sigma_t$ with Z is higher after the solenoid than before. The two contributions of the solenoid to the increase of $\sigma_t$ both become greater when the magnetic field amplitude increases and, as visible on Fig. 3 (right), the first one is the dominant one for high values of $\sigma_{rL}$ while the second one is dominant for smaller values of $\sigma_{rL}$.

### 2.3. Cases retained for injection into the THz linac

Table 2 gathers the cases we studied for injection into the THz linac and displays the corresponding values of $\sigma_{rLm}$ and the achieved $\sigma_t$ at the entrance of the THz linac. The values shown in Table 2 are for a transverse focusing to 100 µm rms, corresponding to an injection into a 300 GHz linac.

Table 2 shows, as expected, that the increase of the field amplitude in the RF-gun from 110 MV/m to 140 MV/m is of great help for decreasing $\sigma_t$ at the THz linac entrance, since the reduction is between 25% and 30% according to the case considered. This is due to the decrease of the space-charge forces and to the stronger bunch compression at the photocathode. It also shows that putting the THz linac closer from the RF-gun allows reducing $\sigma_t$ at the THz linac entrance by a factor between 13% and 20% according to the case considered. The deterioration of $\sigma_t$ induced by the stronger focusing field of the solenoid in this case is more than compensated by the facts that $\sigma_{rLm}$ is lower and that the space-charge forces have less effect due to the shorter traveled distance.

For the case of injection into a 150 GHz linac, we do not perform again the search for $\sigma_{rLm}$, but keep the same value as for the injection into a 300 GHz linac shown in Table 2. Table 3 shows the values achieved for $\sigma_t$ at the THz linac entrance for this case. By comparing Tables 2 and 3, one can see that $\sigma_t$ is slightly lower for the case of injection into a 150 GHz linac. This is explained by the weaker transverse focusing of the bunch required, due to the larger linac radius, which implies a smaller effect of the solenoid magnetic field on the bunch length. However the effect remains small, since it is at maximum of 6%.

In the following section, concerning the acceleration and compression in the THz linac, we will only consider the cases of Table 2 and 3 with $E_{mg}$ = 140 MV/m (underlined in the two tables).

**Table 2:** $\sigma_{rLm}$ and $\sigma_t$ at the entrance of a 300 GHz linac for the studied cases.

| $E_{mg}$ (MV/m) | 110 | 110 | 110 | 110 | 140 | 140 | 140 | 140 |
|---|---|---|---|---|---|---|---|---|
| Q (pC) | 1 | 5 | 1 | 5 | <u>1</u> | <u>5</u> | <u>1</u> | <u>5</u> |
| $Z_{linac}$ (cm) | 85.6 | 88.5 | 60.9 | 62.1 | <u>85.0</u> | <u>89.1</u> | <u>61.0</u> | <u>62.7</u> |
| $\sigma_{rLm}$ (mm) | 0.45 | 0.7 | 0.4 | 0.6 | <u>0.5</u> | <u>0.75</u> | <u>0.45</u> | <u>0.6</u> |
| $\sigma_t$ (fs) | 39.0 | 85.2 | 33.9 | 72.5 | <u>28.7</u> | <u>65.1</u> | <u>24.1</u> | <u>52.4</u> |

**Table 3:** $\sigma_{rLm}$ and $\sigma_t$ at the entrance of a 150 GHz linac for the studied cases.

| $E_{mg}$ (MV/m) | 110 | 110 | 110 | 110 | <u>140</u> | <u>140</u> | <u>140</u> | <u>140</u> |
|---|---|---|---|---|---|---|---|---|
| $Q$ (pC) | 1 | 5 | 1 | 5 | <u>1</u> | <u>5</u> | <u>1</u> | <u>5</u> |
| $Z_{linac}$ (cm) | 86.2 | 88.9 | 60.7 | 62.6 | <u>85.1</u> | <u>88.8</u> | <u>60.9</u> | <u>62.6</u> |
| $\sigma_{rLm}$ (mm) | 0.45 | 0.7 | 0.4 | 0.6 | <u>0.5</u> | <u>0.75</u> | <u>0.45</u> | <u>0.6</u> |
| $\sigma_t$ (fs) | 37.3 | 81.8 | 32.2 | 69.9 | <u>27.0</u> | <u>61.9</u> | <u>22.6</u> | <u>50.4</u> |

## 3. Into the THz linac

### 3.1. Assumptions and conditions

To simulate the beam dynamics into the $TM_{01}$ mode of the THz linac, we used an ASTRA model based on the superposition of two standing waves. More details on the assumptions, possibilities and limitations of this model can be found in [28]. We chose to study the cases of accelerating frequencies $f$ = 300 GHz and $f$ = 150 GHz in the THz linac and two different interaction lengths $L$ between the electron bunch and the THz pulse for these two cases, namely 6.4 cm and 12.8 cm. We fix the field amplitude $E_{ml}$ in the THz linac such that the maximum output energy of the bunch as a function of its injection phase into the linac is around 20 MeV. This is a function of the field amplitude $E_{mg}$ in the RF-gun and results in: $E_{ml}$ = 220 MV/m for $E_{mg}$ = 140 MV/m and $L$ = 6.4 cm; $E_{ml}$ = 115 MV/m for $E_{mg}$ = 140 MV/m and $L$ = 12.8 cm.

We simulate in this section the possibility to use the THz linac to simultaneously accelerate the bunch up to energies between 15 and 20 MeV and to compress it to rms lengths of the single femtosecond order or below by velocity bunching. One of the key features is that in these conditions the bunch length will quickly increase after passing the point of maximal compression, because even at 15-20 MeV the velocity differences between the different electrons of the bunch are significant since the bunch is very short. It is therefore not worth to search for a point of maximal compression right at the THz linac exit, since some drift space is needed after it to transversely focus the bunch to the potential ICS point. In this section, we assume that a drift space of around 30 cm is needed and we have therefore performed ballistic bunching [1] with the THz linac, by varying the dephasing between the accelerating field and the bunch, to put the point of maximal compression ($\equiv$ longitudinal focal point) at this distance. Table 4 gathers the results obtained for all the cases selected at the THz linac entrance in Tables 2 and 3. These results will be analyzed in the following subsections.

**Table 4:** Beam properties obtained 30 cm after the THz linac exit for the cases selected in Tables 2 and 3.

| $Q$ (pC)/$f$ (GHz) | | 1/300 | 5/300 | 1/150 | 5/150 | 1/300 | 5/300 | 1/150 | 5/150 |
|---|---|---|---|---|---|---|---|---|---|
| $E_{ml}$ (MV/m)/ $L$ (cm) | | 220/ 6.4 | 220/ 6.4 | 220/ 6.4 | 220/ 6.4 | 115/ 12.8 | 115/ 12.8 | 115/ 12.8 | 115/ 12.8 |
| Energy (MeV)/ $\sigma_E$ (keV)/ $\sigma_t$ (fs) | $Z_{linac} \approx$ 85 cm | 16.6/ 349/ 2.4 | 16.8/ 824/ 11.3 | 14.0/ 233/ 1.0 | 14.1/ 516/ 6.0 | 17.4/ 234/ 2.2 | 17.6/ 590/ 7.9 | <u>14.9/ 197/ 0.97</u> | 14.8/ 430/ 5.1 |
| | $Z_{linac} \approx$ 60 cm | 16.5/ 294/ 1.7 | 16.8/ 676/ 7.3 | **13.7/ 119/ 1.4** | 13.6/ 305/ 4.2 | **17.2/ 212/ 1.6** | 17.5/ 496/ 4.9 | <u>14.6/ 80/ 3.7</u> | 14.1/ 270/ 3.7 |
| $\varepsilon_{x,y}$ ($\pi$.mm.mrad) | $Z_{linac} \approx$ 85 cm | 0.33 | 0.81 | 0.42 | 1.05 | 0.31 | 0.93 | 0.41 | 0.95 |
| | $Z_{linac} \approx$ 60 cm | 0.35 | 0.78 | 0.32 | 1.24 | 0.36 | 0.84 | 0.28 | 1.38 |

### 3.2. Influence of the THz linac frequency

The most visible influence of the THz linac frequency in Table 4 is that a decrease from 300 GHz to 150 GHz is accompanied by a significant decrease of the energy spread and of the bunch length, except for the particular case written in italic for which a strong increase of the bunch length is observed. This case will be explained in Sec. 3.4.

These decreases are both explained by the fact that a bunch with the same initial length covers a smaller fraction of the accelerating field wavelength when the THz frequency decreases. First, less energy spread is therefore induced by the off-crest acceleration, necessary to compress the bunch by velocity bunching, in the THz linac. Second, less non-linearities are induced during the compression by velocity bunching, because the field curvature seen by the bunch decreases, resulting in a shorter bunch length. The latter effect is clearly visible in Fig. 4 (left), where the bunch longitudinal phase-spaces at the longitudinal focal point are depicted for two cases with the same bunch at the THz linac entrance, differing only by the simulated linac frequency (150 GHz and 300 GHz).

A drawback of the decrease of the linac frequency from 300 GHz to 150 GHz is that it implies a decrease of the bunch final kinetic energy (see Table 4). For the cases shown in Table 4, the decrease is between 15% and 20%. This is explained by the fact that after the initial off-crest injection into the THz linac, necessary to compress the bunch by velocity bunching, the phase-slippage of the bunch respective to the THz field will be slower for lower frequencies. As a result the bunch will move towards less accelerating field values during its path in the THz linac than at higher frequencies (see Fig. 4 (right)) and therefore gain less energy.

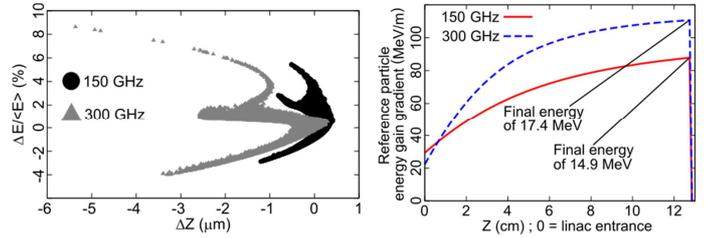

**Fig. 4.** Longitudinal phase-space at the longitudinal focal point (left) and energy gain gradient of the reference particle in the THz linac (right). Conditions: $Q$ = 1 pC; $\sigma_{rL}$ = 0.5 mm; $E_{mg}$ = 140 MV/m; $Z_{linac}$ = 85 cm; $L$ = 12.8 cm; $E_{ml}$ = 115 MV/m.

### 3.3. Influence of the interaction length between the bunch and the accelerating field

The data in Table 4 show that an increase of the interaction length between the electron bunch and the THz field from 6.4 cm to 12.8 cm leads to a decrease of the energy spread and of the bunch length, this latter being more significant for the cases with 5 pC bunch charge. One exception is the particular case written in italic for which a strong increase of the bunch length is observed. This case will be explained in Sec. 3.4.

These decreases are not determined by a reduction of the field curvature seen by the bunch during its path in the THz linac. Indeed, as visible in Fig. 5 (left), with the same bunch at the THz linac entrance, the shape of the longitudinal phase-space at the longitudinal focal point remains the same for the two interaction lengths. The decrease of the energy spread (see Fig. 5 (right)) is actually simply due to the lower field amplitude required when $L$ increases. The decrease of the bunch length is explained by the fact that the longer interaction length and lower field amplitude allows having a slower compression rate of the bunch into the THz linac, which results in a shorter final bunch length. One can recognize here some similarity with the scheme of adiabatic velocity bunching, used in conventional RF accelerating structures, where the bunch compression is performed in two travelling wave accelerating structures instead of one [29].

One can also see in Table 4 that the increase of $L$ leads to an increase of the final bunch kinetic energy, which is positive (let us remind here that the field amplitude in the THz linac has been adjusted for each value of $L$ to provide the same maximal possible output kinetic energy of 20 MeV). As visible in Fig. 5 (bottom), the fraction of the peak field seen by the bunch at the level of the exit of the THz linac for $L = 6.4$ cm is almost the same for the two values of $L$. However, after this point, the bunch will continue to slip in phase respective to the THz field for $L = 12.8$ cm and will experience a higher fraction of the peak field. This more than compensates the fact that before 6.4 cm the fraction of the peak field seen by the bunch is actually lower, and explains the higher final kinetic energy.

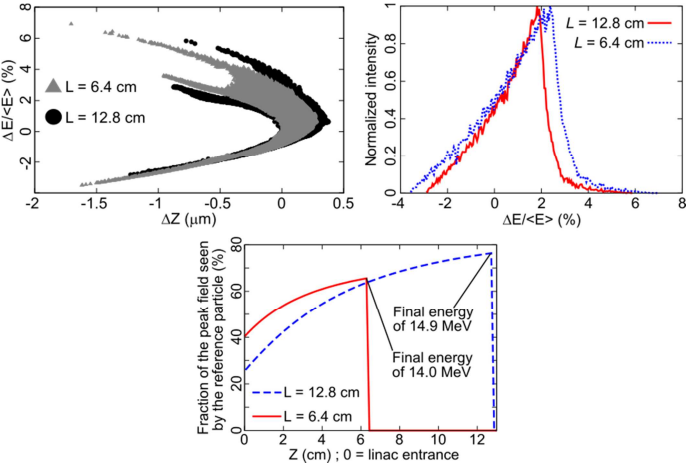

**Fig. 5.** Bunch longitudinal phase-space at the longitudinal focal point (left), bunch energy spectrum at the longitudinal focal point (right) and fraction of the peak field seen by the reference particle in the THz linac (bottom). Conditions: $Q = 1$ pC; $\sigma_{rL} = 0.5$ mm; $E_{mg} = 140$ MV/m; $Z_{linac} = 85$ cm; $f = 150$ GHz; $E_{ml} = 115$ MV/m ($L = 12.8$ cm) and 220 MV/m ($L = 6.4$ cm).

*3.4. Influence of the position of the THz linac entrance*

During our study, we have observed that the position of the THz linac entrance has a complex influence on the achieved bunch properties. The cases we have studied are not yet sufficient to fully characterize this influence. In this section, we will therefore just focus the attention on two important related effects. These two effects both have the same origin, namely the balance between the focusing induced by the solenoid to inject the bunch into the THz linac and the transverse force $F_r$ due to the THz field in the linac. This transverse force can be simply derived from the analytical expression of the TM$_{01}$ mode in a dielectric-loaded circular waveguide [30] as:

$$F_r = \frac{2\pi e E_{ml} f g(r)}{v_{ph}} \left(1 - \frac{v_z v_{ph}}{c^2}\right) \sin(2\pi f t - k_z z + \varphi_0)$$

$$g(r) = \begin{cases} I_1(k_1 r)/k_1 & \text{if } v_{ph} < c \\ r/2 & \text{if } v_{ph} = c \\ J_1(k_2 r)/k_2 & \text{if } v_{ph} > c \end{cases}$$

$v_{ph}$ is the phase velocity of the THz field, $v_z$ the average bunch longitudinal velocity, $r$ the transverse offset, $k_z = 2\pi f/v_{ph}$, $k_1^2 = k_z^2 - k_0^2$, $k_2^2 = -k_1^2$ ($k_0 = 2\pi f/c$) and $J_1$ and $I_1$ are respectively the Bessel and modified Bessel functions of the first kind [31]. One can remark that for a given phase and fixed values of $r$ and $v_z$, $F_r$ changes sign when $v_{ph}$ crosses the value $c^2/v_z$. The important feature to retain is that in our case ($v_{ph} = c$, so $v_{ph} < c^2/v_z$) the transverse force is defocusing at the phases where the bunch is compressed by velocity bunching, becomes zero on the crest of the THz field and then becomes focusing. In our study, we always compress the bunch by velocity bunching. $F_r$ will therefore always counteract the focusing due to the solenoid.

The first effect is that putting the THz linac entrance closer to the cathode, implying a shorter bunch length at the linac entrance (see Tables 2 and 3), does not necessarily imply that the bunch length at the longitudinal focal point will be shorter, as it is visible with the two cases underlined in Table 4. The evolution of the bunch length and transverse dimension from the THz linac entrance to the longitudinal focal point are presented for these two cases in Fig. 6. It is clearly visible on Fig. 6 (right) that for the case with $Z_{linac} \approx 85$ cm after the cathode, the transverse defocusing force in the linac aforementioned is sufficient to balance the focusing due to the solenoid, leading to an almost constant and quite large bunch transverse dimension ($\geq 150$ µm rms). On the other hand, for the case with $Z_{linac} \approx 61$ cm after the cathode, the focusing due to the solenoid is stronger and the transverse defocusing force is insufficient to balance it. As a result, the bunch experiences a transverse focal point (around 30 µm rms) located close before the longitudinal focal point. This leads to a strong increase of the space-charge forces, which limits the achievable bunch length (see Fig. 6 (left)). It explains why the case written in italic in Table 4 does not follow the general tendencies with $f$ and $L$ described in Sec 3.2 and 3.3, and has a longer bunch length than the two cases written in bold in Table 4.

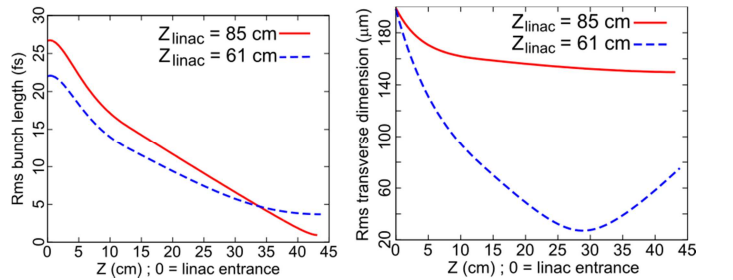

**Fig. 6.** Evolution of the rms bunch length (left) and transverse dimension (right) between the THz linac entrance and the longitudinal focal point for two positions of the linac entrance. Conditions: $Q = 1$ pC; $\sigma_{rL} = 0.5$ mm ($Z_{linac} = 85$ cm) & 0.45 mm ($Z_{linac} = 61$ cm); $E_{mg} = 140$ MV/m; $f = 150$ GHz; $E_{ml} = 115$ MV/m; $L = 12.8$ cm.

The second effect is that this balance between the focusing due to the solenoid and the transverse defocusing force in the THz linac totally determines the bunch transverse properties, dimension and divergence, at the exit of the THz linac since we have assumed that no focusing magnets are placed around the linac. As a result, a case having a short bunch length at the longitudinal focal point (30 cm after the THz linac exit) may turn to be unfavorable because its transverse properties at the linac exit are incompatible with an immediate transverse focusing to the potential ICS point. Fig. 7 shows an example of such a case, where it is visible that at the THz linac exit the transverse dimensions of the bunch are quite small (≈ 45 µm rms) and close to a waist. This makes the transverse focusing of the bunch difficult. As a result, the focusing system to the potential ICS point cannot be placed right at the linac exit in this case, and some drift space has to be added. The consequence is that the potential ICS point, and therefore the longitudinal focal point, has to be placed much farther than 30 cm after the linac exit as assumed in this study, which will penalize the achievable bunch length. An important axis of investigation in future studies will therefore be the possibility to have some control on the transverse bunch properties at the THz linac exit. This would greatly help the final focusing up to the potential ICS point. This could be done by a fine-tuning of the strength of the solenoid after the RF-gun and/or of the position of the THz linac entrance and/or by placing tunable magnets around it.

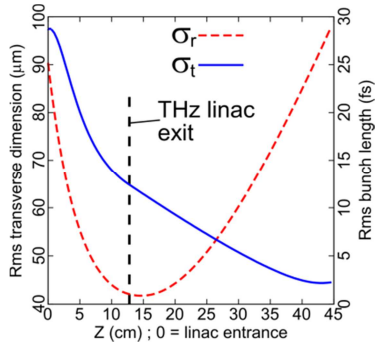

Fig. 7. Evolution of the rms bunch length $\sigma_t$ and transverse dimension $\sigma_r$ between the THz linac entrance and the longitudinal focal point. Conditions: $Q$ = 1 pC; $\sigma_{rL}$ = 0.5 mm; $E_{mg}$ = 140 MV/m; $f$ = 300 GHz; $E_{ml}$ = 115 MV/m; $L$ = 12.8 cm; $Z_{linac}$ = 85 cm.

### 4. Focusing to the potential ICS point

As explained at the beginning of Sec. 3.1, the bunch length will quickly increase after passing the point of maximal compression. When simulating the final focusing up to the potential ICS point, we will therefore take special care that the longitudinal and transverse focal points coincide together to obtain the maximal charge density at this point. We will consider as transverse focusing device a triplet of quadrupole electromagnets, for which we fix the magnetic length of each magnets to be 5 cm and the spacing between each magnet center to be 8 cm. We will first present the best case we obtained for $Q$ = 1 pC. Fig. 8 shows the evolution of the bunch length and transverse dimension from the cathode up to the potential ICS point and a zoom around this point. Fig. 9 shows the bunch longitudinal phase-space and transverse profile at the potential ICS point and Table 5 gathers the bunch properties at this point.

Fig. 8 (right) demonstrates that the ballistic bunching imprinted by the THz linac can be precisely tuned in order to match the longitudinal and transverse focal points at the potential ICS point. Then, Fig. 9 shows that the bunch transverse profile at the ICS point is almost totally Gaussian with a faint halo. This is important since to optimize the ICS the bunch transverse profile has to be matched to the one of the laser, which would typically be Gaussian. Fig. 9 also shows that the bunch longitudinal phase-space exhibits no spikes of current. Finally, the properties presented in Table 5 fulfil, or almost fulfil, the requirements for the AXSIS project presented in Table 1, showing that the THz linac is a very promising option to fulfil the AXSIS requirements. These results also demonstrate that a dedicated hybrid layout mixing an S-band gun and a THz linac has an excellent potential to be a compact X-ray source based on ICS, delivering fs and potentially sub-fs pulses.

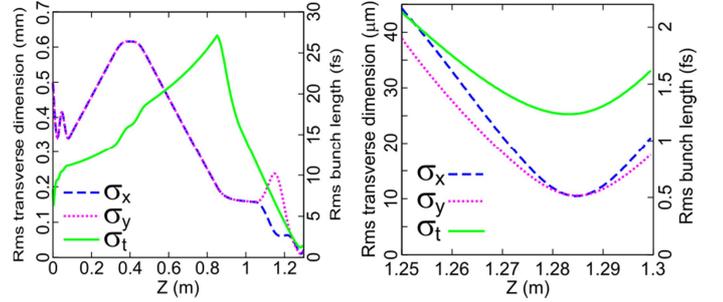

Fig. 8. Evolution of the rms bunch length $\sigma_t$ and transverse dimensions ($\sigma_x$;$\sigma_y$) from the cathode to the ICS point (left) and zoom around this point (right). Conditions: $Q$ = 1 pC; $\sigma_{rL}$ = 0.5 mm; $E_{mg}$ = 140 MV/m; $Z_{linac}$ = 85 cm; $f$ = 150 GHz; $E_{ml}$ = 115 MV/m; $L$ = 12.8 cm; Center of first quadrupole 10 cm after the linac exit; Gradients: +8.5/-19/+29.375 T/m.

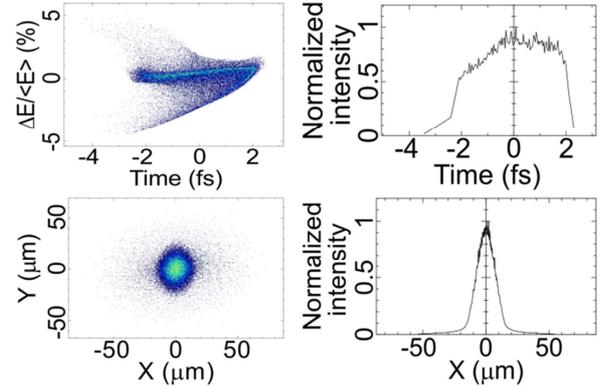

Fig. 9. Bunch longitudinal phase-space (top) and transverse profile (bottom) at the potential ICS point. Conditions: See Fig. 8.

Table 5: Bunch properties at the potential ICS point. Conditions: See Fig. 8.

| Charge | Mean kinetic energy | Rms energy spread | Rms length | Rms transverse dimension (horizontal*vertical) | Rms transverse emittance (horizontal*vertical) |
|---|---|---|---|---|---|
| 1 pC | 14.85 MeV | 153 keV (1.03%) | 1.24 fs | 10.47*10.56 µm² | 0.365*0.284 (π.mm.mrad)² |

Several limitations can be observed in the case shown in Fig. 8, Fig. 9 and Table 5. First, the electron bunch does not yet reach the sub-fs level since $\sigma_t$ = 1.24 fs at the potential ICS point. Then, the transverse emittance is rather high considering the low charge of 1 pC we have used. This is linked to the quite high transverse dimension used for the UV laser pulse generating the bunch, which is required for minimizing the bunch length at the THz linac entrance (see Fig. 3, Table 2 and Table 3). This results in a high thermal emittance value (0.19 π.mm.mrad for $\sigma_{rL}$ = 0.5 mm as for the case shown in Table 5) which is directly proportional to the transverse dimension on the cathode [32]. Finally, the relative energy spread is quite high (around 1% rms). This would be penalizing for a potential X-ray source based on ICS since the natural relative bandwidth of the emitted radiation is increased, among other contributions, by two times this relative energy spread [33]. Another important limitation is that the achieved bunch properties at the ICS point are quickly deteriorated when the bunch charge increases, because of the space-charge forces. As an illustration of this fact, the current status of our work with a 5 pC bunch charge is displayed in Fig. 10, showing the bunch longitudinal phase-space and transverse profile at the potential ICS point, and Table 6, gathering the bunch properties at this point. Numerous potential solutions are currently under investigation to try to overcome these limitations and improve the achieved bunch properties, both at 1 pC and 5 pC. The implementation of these potential solutions will lead to a further optimization of the simulated hybrid layout presented in this paper, which will be presented in a future publication.

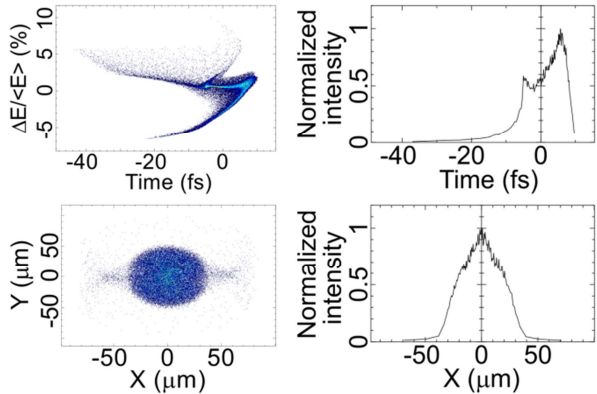

**Fig. 10.** Bunch longitudinal phase-space (top) and transverse profile (bottom) at the potential ICS point. Conditions: $Q$ = 5 pC; $E_{mg}$ = 140 MV/m; $Z_{linac} \approx$ 63 cm; $f$ = 150 GHz; $E_{ml}$ = 115 MV/m; $L$ = 12.8 cm; Center of first quadrupole 21 cm after the linac exit; Gradients: +15/-23/+26.5 T/m.

**Table 6: Bunch properties at the potential ICS point. Conditions: See Fig. 10.**

| Charge | Mean kinetic energy | Rms energy spread | Rms length | Rms transverse dimension (horizontal*vertical) | Rms transverse emittance (horizontal*vertical) |
|---|---|---|---|---|---|
| 5 pC | 15.20 MeV | 293 keV (1.93%) | 7.75 fs | 19.73*25.37 µm² | 1.076*1.176 (π.mm.mrad)² |

## 5. Conclusions

Our simulation study demonstrates, comparing Table 5 to Table 1, that the type of THz linac intended to be used in the AXSIS project is very promising to fulfil the requirements in terms of bunch properties at the ICS point.

Our study also shows that a dedicated hybrid layout mixing an S-band gun and a THz linac of the same type as the one intended to be used in the AXSIS project has an excellent potential to be a compact X-ray source based on ICS, delivering fs and potentially sub-fs pulses. Fig. 11 presents a schematic of the layout corresponding to the case we present in Fig. 8 and Fig. 9, with the distances and relevant parameters and bunch properties at the ICS point.

The study presented in this paper also allowed us identifying several limitations to the scheme we propose: the difficulty to control the transverse bunch properties at the THz linac exit and subsequently the final focusing to the ICS point; the bunch transverse emittance and relative energy spread are rather high; the electron bunch does not yet reach the sub-fs level. Another important limitation is that the achieved bunch properties at the ICS point are quickly deteriorated when the bunch charge increases. Numerous potential solutions are currently under investigation to try to overcome all these limitations and improve the achieved bunch properties. This will lead to a further optimization of the hybrid layout presented in this paper.

One should be careful that in our study we considered only beam dynamics aspects, and we did not yet look at the technical feasibility or mechanical integration of the hybrid scheme we proposed. We plan to investigate these aspects in future studies, especially looking at the requirements in terms of beam diagnostics and steering and in terms of properties of the THz pulse driving the THz linac. The goal will be to identify which distances and/or parameters of the schematic shown in Fig. 11 are too technically challenging or too hard to mechanically integrate in a beamline and therefore to be adapted in future layouts.

The optimization of the hybrid layout introduced in this paper, together with the inclusion of technical limitations, will lead to a future publication.

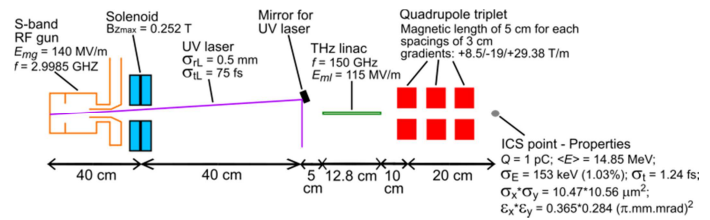

**Fig. 11.** Schematic summarizing the simplified layout we studied in this paper, the parameters we assumed for the simulation of the case shown in Fig. 8 and Fig. 9 and the bunch properties obtained at the potential ICS point.

**Funding**

This work has been supported by the European Research Council under the European Union's Seventh Framework Programme (FP/2007-2013)/ERC Grant Agreement n. 609920.

**Acknowledgments**

The authors would like to thank K. Flöttmann and all the colleagues of the AXSIS collaboration for fruitful discussions.